# Observation of sub-Doppler absorption in the Λ-type three-level Doppler-broadened cesium system

Junmin WANG [§], Yanhua WANG, Shubin YAN, Tao LIU, Tiancai ZHANG

(State Key Laboratory of Quantum Optics and Quantum Optics Devices,

and Institute of Opto-Electronics, Shanxi University,

36 Wucheng Road, Taiyuan, Shanxi 030006, China)

**Abstract**: Thanks to the atomic coherence in coupling laser driven atomic system, sub-Doppler absorption has been observed in Doppler-broadened cesium vapor cell via the Λ-type three-level scheme. The linewidth of the sub-Doppler absorption peak become narrower while the frequency detuning of coupling laser increases. The results are in agreement with the theoretical prediction by G**.** Vemuri *et al*.

## 1. Introduction

Coherence between the energy levels of atomic system attracts more attentions and plays a very important role in recent years. There are many effects due to the atomic coherence, such as amplification without inversion (AWI) [1], electromagnetically induced transparency (EIT) [3], coherent population trapping (CPT) [3], electromagnetically induced absorption (EIA)[4], etc. The atomic coherence also has been used in various applications, for instance, the velocity selected coherent population trapping (VSCPT) method in sub-recoil laser cooling [5], enhancement of the refractive index [6], slow group velocity of optical pulse in ultra-cold atomic cloud [7], and light storage [8]. Actually most of the experiments on AWI, EIT, CPT and EIA are performed in room-temperature atomic vapor.

Around room temperature the resolution of absorption spectrum in atomic vapor is usually dominated by Doppler broadening because of the Maxwell-Boltzman distribution of the atomic velocity. Based on velocity selected optical pumping saturated absorption spectroscopy (SAS) with counter-propagating pump and probe beams from the same laser was developed. The SAS technique can realize sub-Doppler resolution in absorption spectra of the probe. Is there any other way to obtain sub-Doppler resolution absorption? G. Vemuri *et al* [9] theoretically demonstrated the possibility of realizing sub-Doppler resolution in the Ladder-type or Λ-type three-level Doppler-broadened atomic system via the atomic coherence induced by a coupling laser, where the atomic coherence is used to influence inhomogeneous broadening. Following the above theoretical work ref [10] and [11] reported the experimental results performed in rubidium vapor cell.

---

[§]   Corresponding author, email: wwjjmm@sxu.edu.cn

In this article, we experimentally demonstrate sub-Doppler absorption in Doppler-broadened cesium vapor at D2 line via the atomic coherence. The coupling and probe beams are provided by two independent grating-extended-cavity diode lasers with typical output of less than 40mW, compared with intense Ti:Sapphire laser in ref [10] (in which about 600-mW output from Ti:Sapphire laser was used for the coupling beam). A narrower linewidth of the sub-Doppler absorption peak is measured when the coupling detuning increases. When the coupling detuning reachs 812MHz, much larger than the coupling Rabi frequency (about 90MHz in our experiment), the sub-Doppler linewidth of about 6.8MHz which is close to the natural linewidth of cesium D2 line is obtained. It is in good agreement with the simulation based on the theoretical prediction for the Λ-type three-level model in ref [9].

## 2. Doppler-broadened Λ-type three-level cesium system

The relevant energy levels of cesium are schematically shown in Fig.1. Considering the Λ-type three-level system formed by |1>, |2> and |3>, the common upper level |3> is the $6\ ^2P_{3/2}$ excited state, and the lower levels |1> and |2> are the F=3 and F=4 hyperfine states of the $6\ ^2S_{1/2}$ ground state with the splitting of 9.19GHz. The Doppler broadening dominates the linear absorption spectrum. In despite of several hyperfine levels (F'=2, 3, 4 and 5) of the $6\ ^2P_{3/2}$ excited state one cannot distinguish them in the linear absorption spectrum around room temperature because the hyperfine splitting of the $6\ ^2P_{3/2}$ state is well below the full Doppler linewidth (~ 560 MHz). A weak probe laser couples the |1> - |3> transition, and another relatively strong coupling laser couples the |2> - |3> transition with a detuning of $\Delta_C$ and a Rabi frequency of $\Omega_C$. The lifetime of the $6\ ^2P_{3/2}$ excited state is 30ns, so that the total spontaneous decay rate $\Gamma_{31}+\Gamma_{32}$ equals to $2\pi \times 5.3$ MHz.

Let us assume the probe and coupling beams propagates through a atomic vapor cell along a same direction. In this case the two-photon Raman process in this Λ-type three-level system will be near resonance for near all atoms with different velocity class in the interaction region. In dressed state picture the coupling laser splits the upper common level into two dressed states due to the AC Stark effect. When the coupling laser is exactly resonant with the |2> - |3> transition, the probe laser will show the absorption peaks at the location of the two dressed states (the Autler-Townes doublet). A absorption reduction dip can be observed at the location of original |3> state. This is known as EIT [1].

The absorption of the probe laser versus the coupling detuning in this kind of Λ-type three-level system has been theoretically investigated by G. Vemuri *et al* [9]. The absorption characters are modified by the atomic coherence induced by a coupling laser. The linewidth of the two absorption peaks $\Delta v_+$ and $\Delta v_-$ were derived in ref.[9] and were given by

$$\Delta v_{\pm} = \frac{\Gamma_{31} + \Gamma_{32} + 2D}{4}(1 \mp \frac{\Delta_C}{\sqrt{\Delta_C^2 + 4\Omega_C^2}}),$$

here D is the full Doppler lindwidth, which is about 560 MHz for cesium atomic vapor at room temperature. If the coupling detuning $\Delta_C$ is much bigger than the coupling Rabi frequency $\Omega_C$ the second term in the bracket of above formula will close to 1. Thereby the linewidth of the two absorption peaks is clearly asymmetric. When $\Delta_C$ is in blue side the $\Delta v_+$ will be much smaller and maybe smaller than D, in another words, a sub-Doppler resolution absorption peak will be observed. Also the larger blue detuning $\Delta_C$, the narrower linewidth of the sub-Doppler absorption peak.

## 3. Experiment and the results

Schematic diagram of our experimental setup is shown in Fig.2. A commercial Littrow-type grating-extended-cavity diode laser (TuiOptics, DL-100) with output of less than 40mW serves as the coupling laser. Another 850nm laser diode (SDL, 5411-G1) with a home-made Littrow-type grating-extended-cavity is used for the probe laser. Both laser systems operate near the cesium $D_2$ transition (852nm). The linewidth of the two lasers is checked to be about 2 MHz in the time scale of 100ms by heterodyne beat method, compared with 5.3MHz of the natural linewidth of $6\ ^2P_{3/2}$ state. Both lasers' output is shaped to near circular beam by anamorphic prism pairs. Optical isolators are used to avoid the optical feedback to keep both lasers stable operation. Experiments are performed in a 30mm-long cesium vapor cell placed in a magnetic shield tube to eliminate the influence of the earth magnetic field and other background magnetic filed. The coupling and probe beams co-propagate through the cesium vapor cell with orthogonal linear polarization configuration. Two polarization beam splitting cubes with typical extinction ratio of about 40dB are used to separate the pump and probe beams. The coupling beam has a spot size of about 2.4mm with power of 28.3mW at the cell region (estimated the Rabi frequency of the coupling laser $\Omega_C \sim$ 90MHz), while the probe beam spot is about 1.2mm with 14μW. The transmission of the probe is detected by a photodiode and viewed or saved by a digital oscilloscope when the probe laser scans over the $6\ ^2S_{1/2}$ F=3 - $6\ ^2P_{3/2}$ transition. The typical scanning speed is about 10Hz. Roughly in 100ms one scan can be accomplished. The saturation absorption spectrometer (depicted 'SAS' in Fig.2) provides a reference frequency standard.

In absent of the coupling beam, Doppler-broadened absorption curve of the $6\ ^2S_{1/2}$ F = 3 - $6\ ^2P_{3/2}$ transition is obtained. The full Doppler linewidth is about 560MHz, which is calibrated by the hyperfine splitting of $6\ ^2P_{3/2}$ state via saturation absorption spectrum. Peak absorption is about 52%. When the coupling beam exists and is roughly resonant with $6\ ^2S_{1/2}$ F = 4 - $6\ ^2P_{3/2}$ transition the maximum absorption of the probe is about 87%, which is much stronger than the case without the coupling beam. This is due to the population redistribution arised by the optical pumping effect of the coupling laser. A small absorption reduction dip at near zero probe detuning due to EIT is observed and shown in Fig.3(a). Fig.3(b) shows the EIT dip as the detuning of the pumping laser frequency $\Delta_C$ is about 346MHz. The absorption reduction is somewhat small. Exactly here the Λ-type system in cesium is not a simple three-level system because the common upper state $6\ ^2P_{3/2}$ has four hyperfine states F'=2, 3, 4, 5. Actually the coupling laser only connects $6\ ^2S_{1/2}$ F = 4 to $6\ ^2P_{3/2}$ F'= 3, 4, 5 hyperfine states according to the dipole transition selection rule ($\Delta F$ = -1, 0, +1), while the probe laser connects $6\ ^2S_{1/2}$ F = 3 to $6\ ^2P_{3/2}$ F'= 2, 3, 4 states. The atomic coherence only exists between F = 4 and F'=3, 4. The components involved F'=2 and F'= 5 can be regarded as incoherence components.

The signal in Fig.3 looks somewhat distorted. When we change the DC voltage of the pizeo behind the grating of ECDL1 to adjust the coupling detuning this will slightly change the output direction of coupling laser. Tthis point is the common shortcoming in the Littrow-configuration grating ECDL. Although the beam splitting cube used to reject the coupling beam into the photodiode has an extinction ratio of about 40dB a very small portion of the coupling beam (about several μW) still can pass through this cube. At certain points this part maybe reach the photodiode inducing the distortion in Fig.3. This is confirmed in the case of without coupling beam and with a larger coupling detuning (Fig.4). For probe beam actually in our experiment only several ten μW power is needed so we insert a small optical wedge inside the grating-extended-cavity and get the probe beam from reflection part of this wedge to avoid the above mentioned output direction problem.

Fig. 4(a) shows typical absorption spectrum of the probe in case of larger blue detuning ($\Delta_C$ = 812 MHz) of the coupling laser. Here the condition of $\Delta_C \gg \Omega_C \sim 90$ MHz is well fulfilled. Fig.4(b) focuses on the sub-Doppler absorption peak. The linewidth $\Delta v$ = 6.8 MHz is obtained by Lorentz fitting to the experimental curve, and it is close to the natural linewidth of 6 $^2P_{3/2}$ state (5.3MHz). Larger coupling detuning is not attempt in our experiment. According to the model in ref [12], in the case of large coupling detuning the linewidth of the coupling and probe laser will clearly infect the observed sub-Doppler linewidth in experiment. Considering our case, the typical linewidth of our coupling and probe laser is about 2 MHz which is well below the natural linewidth (5.3MHz) of 6 $^2P_{3/2}$ state. If we increase the coupling detuning further more maybe the sub-natural absorption peak can be observed. Actually although the coupling and probe approaches are bit different ref [11] demonstrated a narrow peak (5.1MHz) which is slightly narrower than the natural linewidth (6MHz) in rubidium vapor cell with a 1.18GHz coupling detuning. It is note that the mutual coherence between the coupling and probe lasers also will infect the sub-Doppler peak observed in experiment [13]. If we use only one laser and split a small portion to sever as the probe after two acousto-optical frequency shifters utilized to scan the probe frequency better mutual coherence will be realized. So narrower linewidth can be expected.

The linewidth at different coupling detuning is also measured and is shown in Fig.5 as square points. Using the theoretical results of ref [9] based on the $\Lambda$-type three-level model a simulation for our case is performed. Solid line in Fig.5 denotes the prediction curve of the linewidth of the sub-Doppler absorption peak versus the coupling blue detuning by our simulation. The experimental results are qualitatively in agreement with the theoretical prediction.

In summary, we have experimentally investigated the sub-Doppler absorption in room-temperature Doppler-broadened $\Lambda$-type cesium system via the atomic coherence. The sub-Doppler resolution absorption is realized when the coupling laser operates with a large blue detuning. The sub-Doppler linewidth depending upon the coupling detuning is experimentally measured and is qualitatively in agreement with the simulation. Also the EIT absorption reduction is demonstrated in experiment. A natural linewidth comparable narrow absorption peak via atomic coherence may find its application in frequency stabilization of lasers. It will show better frequency stability than the saturation absorption locking technique in which an attainable resolution is normally much wider than the natural linewidth of hyperfine transition due to various broadening factors.

**Acknowledgments**: The authors are very grateful to Prof. Kunchi PENG for stimulating encouragement. Junmin Wang thanks Hong Chang and Jianming Zhao for useful discussions. This project is financially supported by the National Natural Science Foundation of China (60178006), by the Natural Science Foundation of Shanxi Province (20021030), and by the Research Fund for the Returned Abroad Scholars of Shanxi Province.

**Figure Captions:**

**Fig. 1**   Sketch of the relevant energy levels of Cesium $D_2$ trasition and the Lamda-type three-level scheme. $\Delta_C$ is the detuning of the coupling laser and $\Omega_C$ is the Rabi frequency.

**Fig. 2**   Schematic diagram of the experimental setup. **ECDL:** grating-external-cavity diode laser system;   **AP:** anmorphic prism pairs;   **Isolator:** optical isolator; **λ/2:** half wave plate;   **PBS:** polarization beam splitter cube;   **SAS:** saturated absorption spectrometer;   **PD:** photodiode.

**Fig. 3**   Absorption reduction dip due to EIT at roughly exact resonance (a) and blue detuning (b) of the coupling laser.

**Fig. 4**   Typical absorption curve with a large blue detuning ($\Delta_C$ = 812) of the coupling laser (a). A small absorption peak with sub-Doppler linewidth is observed. While the detuning of the coupling laser $\Delta_C$ increases the sub-Doppler peak moves towards the blue side and the linewidth $\Delta\nu$ decreases. $\Delta\nu$ is obtained by Lorentz fitting to the sub-Doppler peak at different coupling detuning. Fiure 4(b) indicates $\Delta\nu$ = 6.8 MHz corresponding to $\Delta_C$ = 812 MHz.

**Fig. 5**   Linewidth of the sub-Doppler absorption peak versus the frequency detuning of the coupling laser. The square points are experimental results, while the solid line is theoretical prediction.

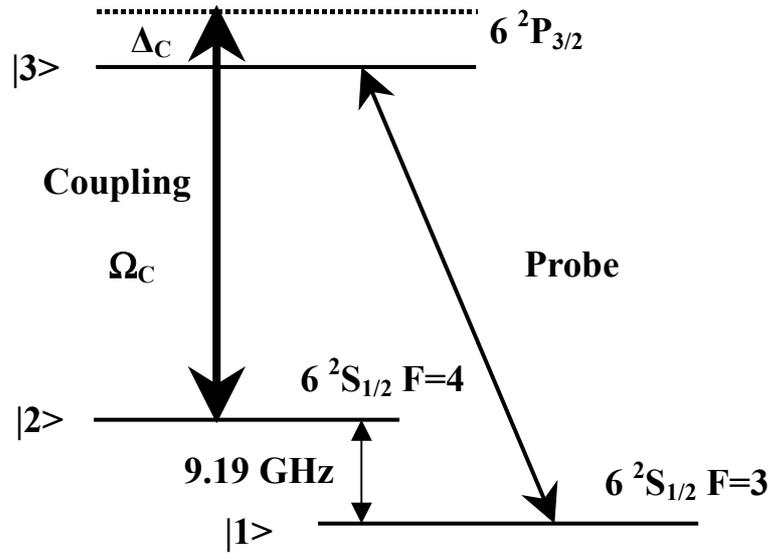

Fig. 1

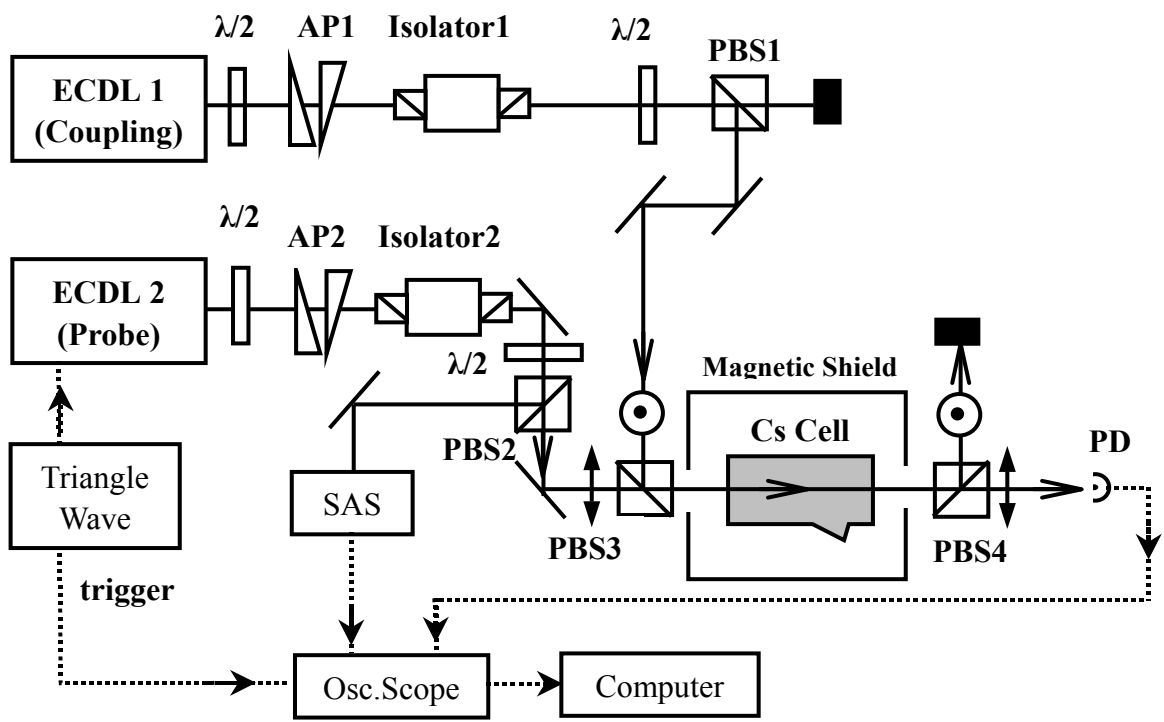

Fig. 2

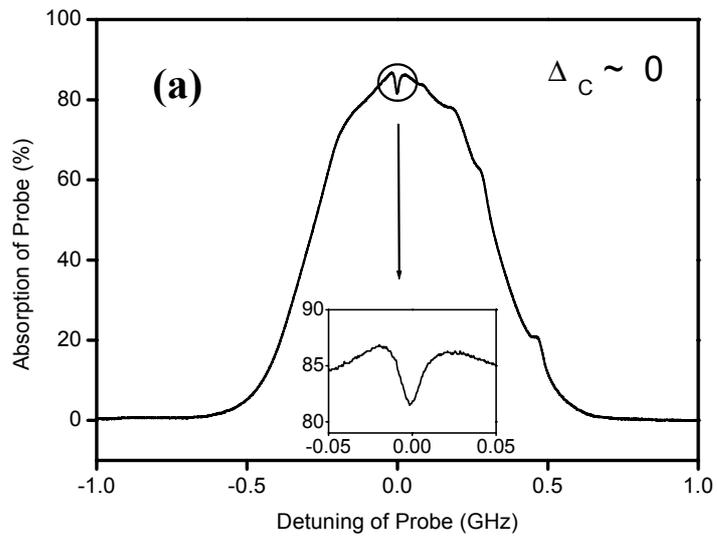

Fig. 3a

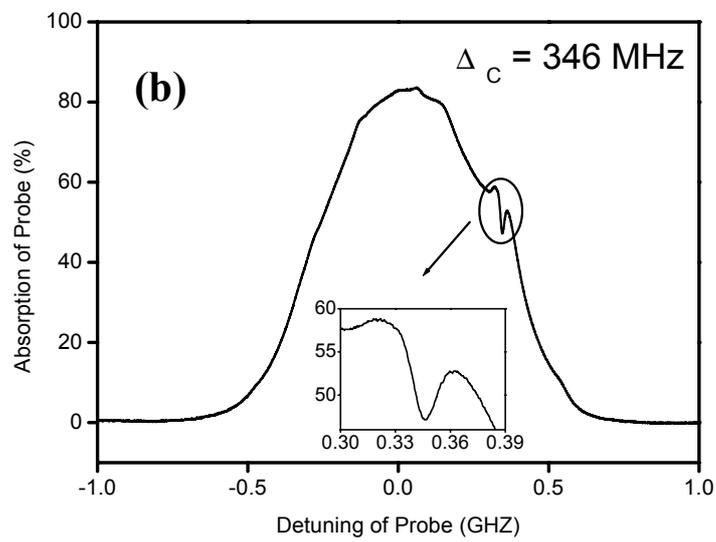

Fig. 3b

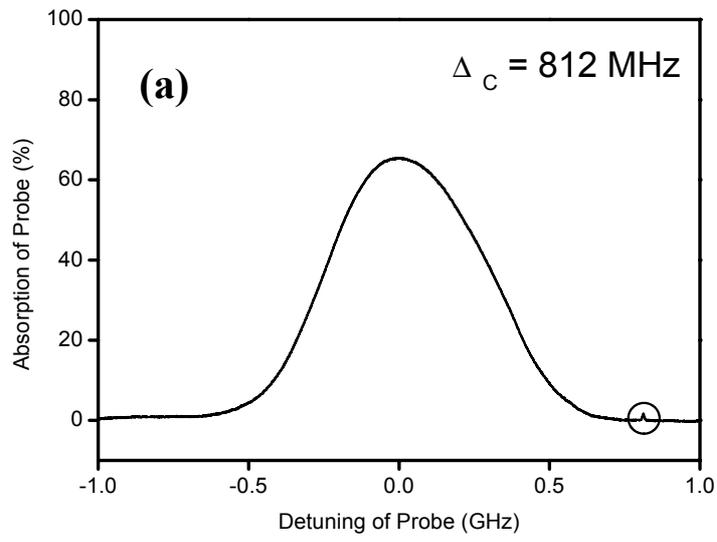

Fig. 4a

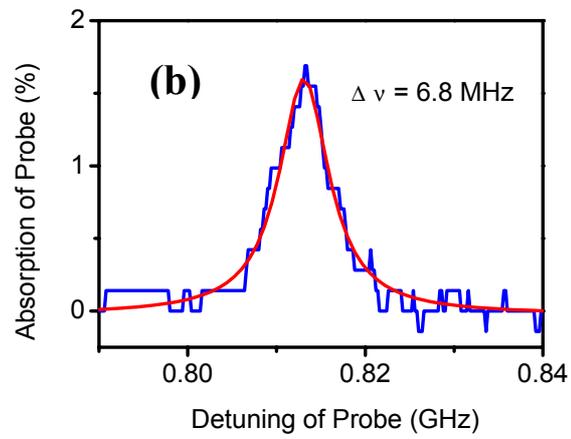

Fig. 4b

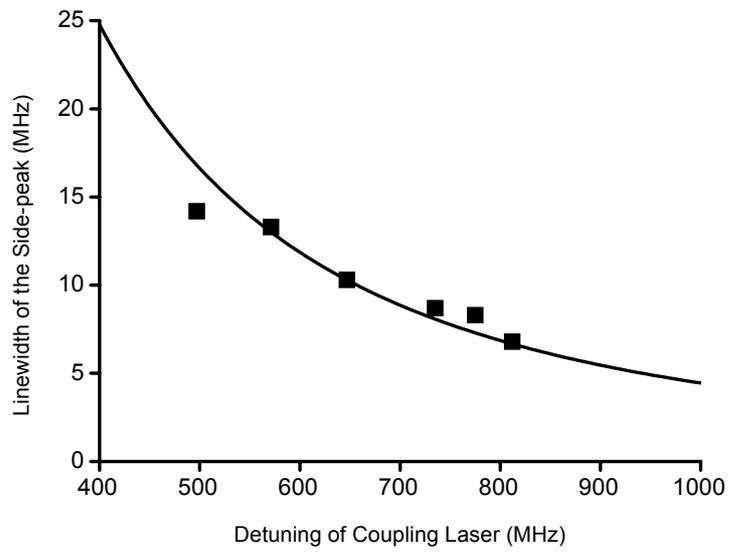

Fig. 5